\documentstyle[seceq,epsf]{ptptex}
\newcommand{\beq}{\begin{equation}}
\newcommand{\eeq}{\end{equation}}
\newcommand{\beqa}{\begin{eqnarray}}
\newcommand{\eeqa}{\end{eqnarray}}
\newcommand{\ba}{\begin{array}}
\newcommand{\ea}{\end{array}}

\title{%        %You can use \\ for explicit line-break
Study of Regular and Irregular States in Generic Systems
}
\author{%       %Use \sc for the family name
Gregor {\sc Veble}, Marko {\sc Robnik} and Junxian {\sc Liu}
\footnote{e-mails: gregor.veble@uni-mb.si, robnik@uni-mb.si, 
jliu@tikva.chem.utoronto.ca}
}

\inst{%         %Affiliation, neglected when [addenda] or [errata]
Center for Applied Mathematics and Theoretical Physics,\\
University of Maribor, Krekova 2, SI--2000 Maribor, Slovenia
}

\abst{%       %this abstract is neglected when [addenda] or [errata]
In this work we semiclassically analyzed the high lying eigenstates 
of a mixed type Hamiltonian system. For the regular states we employ the
Einstein-Brillouin-Keller quantization, while for the chaotic states, 
following the principle of uniform semiclassical condensation, we obtain
a prediction for their wavefunction autocorrelation function.
}

\begin{document}

\maketitle

\section{Introduction}

Classical Hamiltonian systems can range from completely integrable ones, such 
as the Kepler problem and the harmonic oscillator, to fully chaotic systems 
(e.g. Bunimovich stadium, Sinai billiards). Yet the majority of classical 
systems 
fall into neither category but are of the generic, mixed type. The phase space
of such systems is split into areas of quasiperiodic motion on phase-space
tori like in the integrable systems, and into areas where the 
motion is chaotic.

The quantum mechanical properties of classically integrable and classically 
chaotic systems vastly differ in the semiclassical limit $\hbar \to 0$.
The Wigner functions (the quantum mechanical analogues of the classical 
phase space density) of eigenstates of integrable systems are localized on
tori in phase space, and the eigenfunctions in configuration space have an 
ordered structure. On the other hand, the eigenfunctions of fully 
chaotic/ergodic systems
appear random \cite{rf:Berry1977,rf:Voros1979} and their Wigner functions
uniformly cover the whole energy shell in the phase space.

For the mixed type systems the principle of uniform semiclassical condensation
(PUSC, see \citen{rf:Robnik1988,rf:Robnik1998}) states that the Wigner
functions should in this case be localized either on the invariant tori in
the regions of regular motion or should uniformly cover the whole 
chaotic component
of the energy shell in the phase space. This means that states are separated
into regular and irregular ones.

In this work we are interested in geometrical and statistical properties of 
both regular and irregular high lying eigenfunctions. For the regular states
we employ the Einstein-Brillouin-Keller (EBK) quantization, while for
the chaotic states we give an expression for the wavefunction autocorrelation
function based on the geometry of the chaotic component in the classical phase 
space. For more details see reference \cite{rf:Veble1999} and references 
therein. Recently we also performed experiments with microwave cavities on the
same mixed type system as presented here \cite{rf:Veble1999a}, 
while theoretical work on another mixed type system has been done 
by Makino {\it et al.}\cite{rf:Makino1999}

\section{Definitions}

Our model system was a billiard obtained by conformally mapping the unit
circle with the complex quadratic polynomial 
\cite{rf:Robnik1983,rf:Robnik1984},

\beq
 z \to w(z) = z + \lambda z^2,\ w(z)=x + i y.
\eeq 
We chose $\lambda=0.15$, where the classical phase space is roughly
equally divided into components of regular and chaotic motion. The
Poincar\' e surface of section (SOS) was chosen to lie 
on the symmetry axis $y=0$ with coordinate $x$
and the conjugate momentum $p_x$ as the parameters of the surface. The
intersection of the main
chaotic component of our billiard with the SOS is shown in the figure 
\ref{sos}. The coordinate $x$ is taken 
relative to the center of the billiard, while 
$p_x$ is the $x$-component of the unit momentum vector.

The quantum
mechanics of billiards is described by the Helmholtz equation
\beq
(\Delta + k^2) \psi=0,
\eeq
with the Dirichlet boundary conditions, where $k^2=2mE/\hbar^2$.
We limited ourselves to the states with even parity with respect to reflection
across the symmetry line $y=0$. 

For each state we calculated the smoothed 
projection of the Wigner function. The Wigner function of a state 
$\psi({\bf q})$ in the general case of $N$ degrees of freedom
is defined in the full phase space $({\bf q},{\bf p})$ as \cite{rf:Berry1983}
\beq
  W({\bf q},{\bf p})=\frac{1}{(2\pi\hbar)^N}\int d^N 
 {\bf X} \exp(-i {\bf p}\cdot
  {\bf X}/\hbar)\psi^{\dagger}({\bf q}-{\bf X}/2)\psi({\bf q}+{\bf X}/2).
  \label{Wigner}
\eeq
In our case the eigenfunctions $\psi (x,y)$ generate their Wigner transforms 
$W(x,y,p_x,p_y)$ through (\ref{Wigner}), where $N=2$.
In order to compare the Wigner function of a state of our system 
with the classical SOS plot we took its value on the 
symmetry line ($y=0$) and integrated it over $p_y$,
\beq
  \rho_{SOS}(x,p_x)=\int dp_y W(x,y=0,p_x,p_y).
\eeq
The result is
\beq
  \rho_{SOS}(x,p_x)=\frac{1}{2\pi\hbar}\int dX \exp(-i p_x X/\hbar) 
  \psi^{\dagger}(x-X/2, y=0) \psi(x+X/2, y=0).
\eeq
Here we see the reason for considering the even parity states only, because 
$\psi(x,y=0)$ is exactly zero for odd states, and therefore a different 
approach must be used to analyze them.

%As is well known, the Wigner function is not positive definite but exhibits 
%small oscillations that can blur the overall picture. We chose to smooth the
%projection of the Wigner function by a suitable Gaussian. It was chosen 
%narrower
%than the minimum uncertainty Gaussian (in which case the Wigner function 
%becomes the positive definite Husimi distribution) in order not to smooth out
%too many features, but still wide enough to reduce the oscillations.

%The first catalogue of eigenstates and the corresponding smoothed Wigner 
%function projections comprises the first 1000 even states. 
%They were obtained by
%the conformal mapping diagonalization technique (as described in 
%\citen{rf:Robnik1984}). For an analyzis of these states see 
%\citen{rf:Veble1999}.

The catalogue of states studied here consists of $100$
consecutive states starting at the consecutive index of about $2.5\cdot 10^6$.
These states were obtained by the scaling method first introduced by Vergini
and Saraceno \cite{rf:Vergini1995}, 
that enables us to find a few states in the neighbourhood
of a chosen wavenumber $k$. As this is a diagonalizational method no
levels were missed. Almost each level in our small catalogue can be clearly
identified as regular or irregular, an idea proposed already by Percival
\cite{rf:Percival1973}. 
%The only exception in the catalogue is a pair of states lying close 
%together with respect to the mean level spacing, 
%shown in figure
%\ref{pair}, where both of the states are superpositions of a regular ($|\psi_r
%\rangle$) and irregular ($|\psi_i\rangle$) state. These two states
%are close to a degeneracy of two energetically equal but structurally
%different quantized classical objects.
%
%More states of both the regular and irregular type from this catalogue
% with their appropriate analysis will be shown in the next section.

\section{Analysis of states}

In a mixed system the phase space is divided into chaotic and regular 
components. Our work was guided by the principle of uniform semiclassical 
condensation (PUSC, see Robnik 1998), stating that when $\hbar$ tends to $0$ 
the Wigner function of any eigenstate uniformly condenses on an invariant 
object in phase space. This can be either a torus in the regular region or a
whole chaotic component. Each state could thus be
labeled as either regular or irregular (chaotic) in the semiclassical limit.
By looking at the catalogue of states at high (and to some extent even at low)
energies, one can see that this can indeed be done, though there is
still the localization phenomenon present due to the still insufficiently low 
value of the effective Planck's constant.

\subsection{Regular states}
\label{analysis}

We start the analysis by considering 
the regular states. These are the states that can be
attributed to quantized tori within the regular regions. 
For these states we tried to employ the EBK torus quantization. We construct
a wavefunction on the torus as a sum of contributions
\beq
  \psi_j({\bf q})=A_j({\bf q})\exp(i\left[
\frac{1}{\hbar}S_j^{cl}({\bf q})+\phi_j\right])
  \label{semiclassical}
\eeq                   
of different projections $j$ of the torus onto configuration space.
$S_j$ is the classical action with respect to some point on the torus and
$A_j^2$ the classical density of trajectories on this projection.
The phase
of the wavefunction must change by an integer multiple of $2 \pi$
when going around any closed contour of the torus. 
This gives us the quantization conditions
\beq
  I_i= \frac{1}{2\pi} 
   \oint_{\gamma_i}{\bf p}\cdot d{\bf q}=\hbar (n_i+\beta_i/4)
  \label{quantization}
\eeq
where $\gamma_i$ are the irreducible closed contours on the torus and $n_i$ the
torus quantum numbers. The integers $\beta_i$ are Maslov's corrections and 
arise
due to the changes of phase $\phi_j$ at the singularities of projection of 
the torus onto 
configuration space. At each caustic encountered along the contour $\gamma_i$
the wavefunction acquires a negative phase shift of 
$\pi/2$, and shifts by $\pi$ when reflected from a hard 
wall\footnote{If the contour passes the singularity in the contrary
direction to that of the Hamiltonian flow on the torus, the phase shifts are of
the opposite sign}. 
From this consideration it follows that $\beta_i$ counts the number of 
caustics plus twice the number of hard walls encountered along the contour.

The task of finding the semiclassical EBK wavefunctions can be divided into 
two 
parts. The first one is finding the torus with the desired quantum numbers 
$n_i$, the second one being the construction of its appropriate wavefunction
in configuration space. This is not very straightforward since we do not
know the transformation to the torus action-angle variables but can only deal
with the numerically calculated orbits. For more details on this procedure see
reference \cite{rf:Veble1999}.

We show  an example of a regular state in the figure \ref{semi1}. 
We present the exact 
numerical quantum probability density(top left), the probability density 
of its semiclassical approximation (top right), 
the classical density of trajectories on the appropriate torus (bottom left)
and the smoothed projection of the exact Wigner function (bottom right).  
The semiclassical wavefunctions shown are remarkable as they possess
all of the features of their exact counterparts that are larger than the 
appropriate wavelength. Note that for each torus 
there are two characteristic wavelengths since there are two 
quantum numbers associated with it. As it happens in our case, the two 
wavelengths can be of different orders of magnitude.

%The accuracy of the semiclassical energies that we were indeed able to
%reproduce may seem remarkable, since the error is approximately $5$ units of
%energy at the energies around $2\cdot 10^7$. Such accuracy, however, is still
%insufficient to perform short range spectral statistics since the mean level
%spacing in our system is approximately $8$ units of energy. This experience
%is of course in agreement with the proposition and conclusion that the
%semiclassical methods (to the leading order) cannot resolve the energy
%spectra within the vanishing fraction of the mean level spacing, and
%also not the structures of the wavefunctions smaller than de Broglie
%wavelength (Prosen and Robnik 1993a, Robnik and Salasnich 1997).

\subsection{Irregular states}

While for the regular states it was quite straightforward to find their
semiclassical approximations, the nature of irregular states is very much
different. 
%The chaotic component of a system does not possess any obvious
%structure. While Gutzwiller's approach can yield the properties of a quantum
%system by a summation over all periodic orbits of its classical counterpart,
%the relevance of examining individual chaotic states becomes questionable.
These states are very sensitive to small perturbations of the system, so in any
physical system the individual features of the states are lost 
when the effective Planck's constant tends to $0$. The features that are
insensitive
to small perturbations are, however, the statistical properties of spectra and
eigenstates.

One measure of statistical properties of the wavefunctions is the wavefunction
autocorrelation function,
\beq
  C({\bf q}, {\bf x})=\frac{\langle \psi^{\dagger}({\bf q}^{\prime}-{\bf x}/2)
  \psi({\bf q}^{\prime}+{\bf x}/2)
  \rangle_{{\bf q}^{\prime}\in \epsilon({\bf q})}}{\langle\psi^{\dagger}
  ({\bf q^{\prime}})
  \psi({\bf q}^{\prime})\rangle_{{\bf q}^{\prime}\in \epsilon({\bf q})}}.
  \label{autocorrelation}
\eeq
The area of averaging $\epsilon({\bf q})$ close to the point ${\bf q}$ 
should be taken such that its linear size is many wavelengths across, however
small enough that the local properties of classical mechanics within it are
largely uniform.

If one takes the Fourier transform of the Wigner function (\ref{Wigner}),
it is easy to show that
\beq
  \int W({\bf q},{\bf p}) \exp(i {\bf p}\cdot {\bf x}/\hbar) d^{N} {\bf p}=
  \psi^{\dagger}({\bf q}-{\bf x}/2)\psi({\bf q}+{\bf x}/2). 
  \label{fourier} 
\eeq                     
By knowing the Wigner function of an eigenstate,
it is then possible to use this result to calculate
its autocorrelation function.
                                     
According
to the principle of uniform semiclassical condensation, the Wigner
function of any chaotic state should uniformly condense on the whole chaotic
component when the effective $\hbar$ tends to $0$. 
Let us limit ourselves only to the cases of the Hamiltonians with an
isotropic dependence upon ${\bf p}$.
We can write the semiclassical Wigner function as
\beq
  W_{{\cal D}_i}
  ({\bf q},{\bf p})=\alpha \delta(E-H({\bf q},p))
 \chi_{{\cal D}_i} ({\bf q},{\bf p})
\eeq                                
where $\chi_{{\cal D}_i}$ 
denotes the characteristic function of the chaotic component 
and $\alpha$ is the normalization
constant.  

If we write the characteristic function in two degrees of freedom
as a Fourier series of the polar angle $\phi_p$ of the momentum vector (since
the absolute value of ${\bf p}$ is constant at a given energy and point 
${\bf q}$),
\beq
  \chi_{{\cal D}_i}({\bf q},{\bf p})=\sum_{m=-\infty}^{\infty}
  \kappa_m^{{\cal D}_i}({\bf q})
   \exp(i m \phi_p),
\eeq 
it is quite straightforward to show by using the above expressions and 
the integral representations of the Bessel functions that
\beq
 C_{{\cal D}_i}({\bf q}, {\bf x})=\frac{\langle \sum_{m=-\infty}^{\infty}
 \kappa_m^{{\cal D}_i}({\bf q^{\prime}}) 
 i^m J_m(p({\bf q^\prime}) r/\hbar)\exp(i m 
 \phi_x)
 \rangle_{{\bf q}^{\prime}
 \in \epsilon({\bf q})}}{\langle
 \kappa_0^{{\cal D}_i}({\bf q^{\prime}}) 
 \rangle_{{\bf q}^{\prime}\in \epsilon({\bf q})}},
 \label{predic}
\eeq
where $\phi_x$ is the polar angle of the vector ${\bf x}$.

%If the area of averaging is large enough so
%that the variations in the classical phase space picture become important, the
%above procedure is still as long as the value of the 
%momentum does not change
%appreciably (as is the case in our billiard system, where between the
%bounces the 
%momentum remains constant). One can then imagine the large averaging area as
%being cut
%into smaller ones within which the above assumptions still hold true.

We compare the autocorrelation function of a chaotic state with the 
semiclassical prediction \ref{predic}
in figure \ref{chaos1}. 
The averaging area $\epsilon({\bf q})$ 
was taken as a circle of radius $0.2$ around the point $(x,y)=(0.65,0)$.
It was taken the same for both the semiclassical 
prediction and for the numerical results. The averaging 
radius was taken quite large in 
order to reduce the localization properties of the wavefunctions, which are 
still apparent at the values of effective $\hbar$ that we were able to obtain. 
But this radius still has to be taken small enough in order not to completely
smooth out the classical dynamics. The agreement with the semiclassical 
prediction is quite good. It clearly 
deviates from 
the Berry's prediciton\cite{rf:Berry1977} for fully ergodic systems and tends 
towards our semiclassical result.

\section{Conclusion}

In this work we were able to find semiclassical wavefunctions of the regular
eigenstates, while for the chaotic states we 
gave a prediction for the wavefunction 
autocorrelation function, both of which match nicely with their exact
counterparts. What needs to be studied further is the localization
of the chaotic states on only parts of the whole chaotic component 
when not yet in the strict semiclassical limit,
which gives rise to deviations from semiclassical predictions.

\section*{Acknowledgements}
We thank Dr. Toma\v z Prosen for assistance and advice with some computer
programs. This work was supported by the Ministry of Science and Technology of
the Republic of Slovenia and by the Rector's Fund of the university of
Maribor.

\clearpage

\begin{figure}[t]
\caption{The SOS section of the main chaotic component of the $\lambda=0.15$
billiard.}
\label{sos}
\end{figure}

\begin{figure}[t]
\caption{The probability density for a 
 regular state with $k^2=20421387.1741$ (top left), its
 semiclassical approximation with $k_{sc}^2=20421385.96$ (top right), 
 with the quantum numbers on the torus being $n_1=8648$ and $n_2=1$ 
 (16 equally spaced contours from $0$ to the maximum value in both cases).
 In the bottom row we show the classical density 
with 20 contours from 0 to the maximum of the appropriate torus 
 (left) and the
 smoothed projection of the exact Wigner function (right), with ten contours
 from $0$ to the maximum value.}\label{semi1}
\end{figure}

\begin{figure}[t]
\caption{In the top row we show the probability density 
 for the chaotic state with $k^2=20421262.6667$
 (left, 8 contours), the circle denoting the region of averaging,
 with the smoothed projection of its Wigner function (right, 10 contours). 
In the bottom row we plot the wavefunction autocorrelation function
(averaged as explained in text)
in the $x$ (left) and $y$ (right) directions (full), compared to the 
semiclassical prediction (dashed) and the fully ergodic prediction (dotted).}
\label{chaos1}
\end{figure}

\end{document}